\documentstyle[preprint,aps]{revtex}
\begin{document}
\title{Effects of in-medium vector meson masses on low-mass dileptons 
from SPS heavy-ion collisions}
\bigskip
\author{G. Q. Li$^a$, C. M. Ko$^a$, and G. E. Brown$^b$}
\address{$^a$Cyclotron Institute and Physics Department,\\
Texas A\&M University, College Station, Texas 77843, USA\\
$^b$Department of Physics, State University of New York,
Stony Brook, NY 11794, USA}
 
\maketitle
 
\begin{abstract}
Using a relativistic transport model to describe the expansion of the
fire-cylinder formed in the initial stage of heavy-ion collisions
at SPS/CERN energies,
we study the production of dileptons with mass below about 1 GeV
from these collisions.  The initial hadron abundance and their momentum 
distributions in the fire-cylinder 
are determined by following the general features of the results
from microscopic models 
based on the string dynamics and further requiring that 
the final proton and pion spectra and rapidity
distributions are in agreement with available experimental
data.  For dilepton production, we
include the Dalitz decay of $\pi ^0$, $\eta$, $\eta^\prime$, $\omega$
and $a_1$ mesons, the direct decay of primary 
$\rho ^0$, $\omega$ and $\phi$ mesons, and the
pion-pion annihilation that proceeds through the $\rho^0$ meson,
the pion-rho annihilation that proceeds through the $a_1$ meson, 
and the kaon-antikaon annihilation that proceeds through the $\phi$ meson.
We find that the modification of vector meson
properties, especially the decrease of their mass due to the partial
restoration of chiral symmetry, in hot and dense hadronic
matter,  provides a quantitative explanation
of the recently observed enhancement of low-mass dileptons
by the CERES collaboration in central S+Au collisions and by the HELIOS-3
collaboration in central S+W collisions.
\end{abstract}
 
 
 
\section{Introduction}
 
The study of the properties of vector mesons in hot dense matter formed
in heavy-ion collisions is one of the most exciting problems in nuclear 
physics. Theoretically, lattice QCD simulations should provide us with 
the most reliable information about the
temperature and/or density dependence of vector meson properties.
Currently, these simulations can be carried out only for finite temperature 
systems with zero baryon chemical potential, using the quenched approximation.
From the hadronic correlation functions at large spatial separations 
the screening masses of hadrons have been obtained in the lattice QCD 
\cite{detar87,hash93,boyd95}. In a recent paper, Boyd {\it et al} \cite{boyd95}
have showed that, up to about 0.92T$_c$, there is no significant change
of the rho-meson screening mass. They have also found that the quark condensate
does not change very much until one is extremely close to the critical
temperature. This is in contrast with the result from 
chiral perturbation theory which shows a decrease in the quark condensate
\cite{gasser}.
 
Using effective lagrangians that include vector and axial-vector mesons
as massive Yang-Mills fields of the chiral symmetry, Song \cite{song} has
also found that the rho meson mass at finite temperature remains unchanged 
at the leading order of the temperature. On the other hand, Pisarski
\cite{pisarski} has shown using the gauged linear sigma model that the 
rho meson mass increases while the $a_1$ meson mass decreases with 
temperature, and they become degenerate when chiral symmetry is restored.
 
To the contrary, based on the restoration of scale invariance of QCD
for low momentum scales, Brown and Rho \cite{brown91} and
Adami and Brown \cite{adami93} have shown 
that the mass of non-strange vector mesons should be reduced in dense matter.
This is supported by studies using the QCD
sum-rule approach \cite{hatsu92,asak93}, the effective
hadronic model invoking $\bar N N$
polarization \cite{jean94,hatsu94,song95}, and the quark-meson
coupling model \cite{thomas94}.
 
Experimentally, the properties of vector mesons in nuclear medium
can be investigated
by measuring the dileptons produced from heavy ion collisions.
Since dileptons are not subject to the
strong final-state interactions associated with hadronic observables,
they are the most promising probe of
the properties of hot dense matter formed in the initial stage of
high energy heavy ion collisions.
Dileptons have been proposed as useful observables for studying 
the medium modification of the pion dispersion relation at both
finite density and temperature \cite{gale87,xia88}, the in-medium
properties of vector mesons \cite{pis82,koch92,kar93,hat93,li95a}, 
and the phase transition from hadrons to the quark-gluon plasma
\cite{fein76,shur78,chin82,domo83,mcl85,kaja86,red87,xia89,cley91,rus92,asa93,asa94}.
Indeed, our recent study using the relativistic transport model
has shown that in heavy ion collisions
at SIS/GSI energies significant differences exist between
the dilepton spectra with and without medium
modifications of the mass and width of vector mesons \cite{li95a}.
In particular, we have found, using the in-medium rho meson mass from
the QCD sum-rule calculation, that the rho peak from pion-pion
annihilation shifts to around 550 MeV, and its height increases by
about a factor of four.
A calculation similar to that of Ref. \cite{li95a} has been carried out
in the relativistic quantum molecular dynamics (RQMD)
\cite{hoff94}, and a shift in the rho meson peak in the dilepton
invariant mass spectrum is also seen in heavy ion collisions at AGS/BNL
energies.
 
Dileptons have already been measured at Bevalac/LBL by the DLS
collaboration \cite{dls88} in heavy-ion collisions at incident energies
around 1 GeV/nucleon.  Theoretical studies have shown that
the observed dileptons with invariant masses above about 450 MeV
are mainly from pion-pion annihilation \cite{xiong90,wolf93}.
Unfortunately, statistics are not good enough in the Bevalac experiments
to give definite information on the in-medium vector meson properties.
However, similar experiments with vastly improved statistics
have been planned at SIS/GSI by the HADES collaboration
\cite{hades94}.
 
For heavy ion collisions at the SPS/CERN energies, hot and dense
matter is also formed in the initial stage of the collisions.  We expect
that medium effects will lead to a shift in vector meson peaks
in the dilepton invariant mass spectra from these collisions as well.
Experiments from both the HELIOS-3
\cite{helios95} and the CERES \cite{ceres95} collaboration have shown
that there is an excess of dileptons over those known and expected
sources which cannot be explained by uncertainties and errors of the
normalization procedures \cite{tserruya}.
In particular, in the CERES experiment on
central S+Au collisions at 200 GeV/nucleon, a significant enhancement
of dileptons with invariant masses between 250 MeV to 1 GeV over that
from the proton-nucleus collision has indeed been found.
 
In a recent Letter \cite{li95b}, using the relativistic transport model for
the expansion of the fire-cylinder formed in the initial stage of
S+Au collisions, we have shown that the enhancement
of low-mass dileptons can be quantitatively explained by
the decrease of vector meson masses in hot and dense hadronic matter.
In this paper we will expand and improve upon Ref. \cite{li95b}. We
will give the details of our model and the dynamical evolution of
the fire-cylinder. The $\eta$, $\eta^\prime$, $\phi$, and $a_1$ mesons
that have been omitted
in Ref. \cite{li95b} will be included in our model. The medium modification
of the  $\eta$ and $\eta^\prime$ mesons are obtained based on their 
light-quark content in free space, neglecting possible medium modifications
of their mixing, while $a_1$ mesons are introduced based on the 
Weinberg sum rules, which relate the $a_1$ meson mass to that of its 
chiral partner, the rho meson \cite{wein67,kapu94}. These additional
degrees of freedom are useful to reduce the initial temperature and
pion chemical potential introduced in Ref. \cite{li95b}.
For a complete comparison with the experimental
data for dileptons with invariant mass below 1 GeV,
we will also evaluate the Dalitz decay of
$\pi^0$, $\eta$, $\eta^\prime$, $\omega$ and $a_1$
mesons, some of which contribute significantly to dileptons
with mass below 2$m_\pi$. We will also study the sensitivity
of our results with respect to some of the parameters of our model,
e.g., the initial abundance of mesons and their rapidity distributions.
In Section II, we present the extended
Walecka model in which we couple the light quarks of relevant hadrons
to the scalar and vector fields in order to provide
a consistent way to treat the
medium modification of hadrons in the relativistic transport model.
In Section III, we present the details of the relativistic transport
model and the dynamical evolution of the fire-cylinder.
We will also compare our results for particle
spectra and rapidity distributions to the experimental data. The formalism
for evaluating dilepton emission from the expanding fire-cylinder will
be given in Section IV, together with results, discussions and
comparisons with other theoretical calculations as well as with
the experimental data from both the CERES and the HELIOS-3
collaboration. Section V is devoted to some general discussions.
The paper ends with a brief summary in Section  VI.
 
\section{The extended Walecka model}
 
To study consistently the effects of dropping vector meson masses
on the dilepton spectrum in heavy-ion collisions, we need a model for the
in-medium vector meson masses that can be incorporated into the
relativistic transport model to describe the dynamics
of heavy-ion collisions. For this purpose we extend the Walecka model
\cite{qhd86}
from the coupling of nucleons to the scalar and vector fields to the
coupling of light quarks to the scalar and vector fields,
using the ideas of the meson-quark coupling model \cite{thomas94}
and the constituent quark model.
Recently, Gelmini and Ritzi \cite{gr95}, Brown and Rho \cite{br95},
and Furnstahl {\it et al.} \cite{furn95}
have been able to obtain the Walecka mean field model from a 
chiral Lagrangian. Although the latter work appears to be
quite different from the former two, the anomalous dimension of
the loop correction is chosen in \cite{furn95}
to effectively linearize the Lagrangian as for 
Gelmini and Ritzi \cite{gr95} and Brown and Rho \cite{br95}.
We find that these works give us a firm basis for our formulation.
 
We consider a system of baryons (we take here the nucleon as
an example), pseudoscalar mesons ($\pi$ and $\eta$ mesons),
vector mesons (rho and omega mesons), and
the axial-vector meson ($a_1$) at a temperature $T$ and a baryon density
$\rho _B$. The nucleons couple to both the scalar and the vector field,
while the nonstrange mesons other than the pion
couple only to the scalar field, since in the quark
model they are made of a quark and an antiquark
which couple oppositely to the vector field. 
In the mean-field approximation, the thermodynamical potential 
is then given by \cite{furnstahl}
\begin{eqnarray}
\Omega(T,\mu_B)&=&\frac{1}{2}(m_S^2\langle S\rangle^2-m_V^2\langle
V\rangle_0^2)
-T\{{4\over (2\pi )^3}\int d{\bf k}\ln\Big[\exp(-(E^*_N-\mu _B)/T)+1
\Big]\nonumber\\
&+&{4\over (2\pi )^3}\ln\Big[\exp(-(E^*_N+\mu _B)/T)+1\Big]
+{3\over (2\pi )^3}\int d{\bf k}\ln\Big[\exp(-E_\pi /T)-1\Big]\nonumber\\
&+&{1\over (2\pi )^3}\int d{\bf k}\ln\Big[\exp(-E_\eta^* /T)-1\Big]
+{9\over (2\pi )^3}\int d{\bf k}\ln\Big[\exp(-E_\rho ^*/T)-1\Big]\nonumber\\
&+&{3\over (2\pi )^3}\int d{\bf k}\ln\Big[\exp(-E_\omega ^*/T)-1\Big]
+{9\over (2\pi )^3}\int d{\bf k}\ln\Big[\exp(-E_{a_1}^*/T)-1\Big]\},
\end{eqnarray}
where $\langle S\rangle$ and $\langle V_0\rangle$ are the 
scalar and vector mean fields, 
respectively, and the baryon chemical potential 
$\mu_B$ is determined by the baryon density
\begin{eqnarray}
\rho _B =-\frac{\partial\Omega}{\partial\mu_B}=
{4\over (2\pi )^2} \int d{\bf k}\Big[{1\over \exp((E^*_N-\mu _B)/T)
+1}-{1\over \exp((E^*_N+\mu _B)/T)+1}\Big].
\end{eqnarray}
In the above expressions, we have $E_N^*=\sqrt {{\bf k}^2+m_N^{*2}}$
and similar expressions for $E_\eta^*$, $E_\rho^*$, 
$E_\omega^*$ and $E_{a_1}^*$.
Since the medium modification of pions is neglected
in this study, we have $E_\pi =\sqrt {{\bf k}^2+m_\pi^2}$. 
 
The scalar mean field is determined self-consistently
from $\partial\Omega/\partial \langle S\rangle =0$, i.e.,
\begin{eqnarray}
m_S^2\langle S \rangle &=&{4g_S\over (2\pi )^3}\int d{\bf k} {m_N^*\over
E^*_N}\Big[{1\over \exp ((E^*_N-\mu _B)/T)+1}
+{1\over \exp ((E^*_N+\mu _B)/T)+1}\Big]\nonumber\\
&+&{0.45g_S\over (2\pi )^3}\int d{\bf k} {m_\eta^*\over E_\eta ^*}
{1\over \exp (E_\eta ^*/T)-1}+
{6g_S\over (2\pi )^3}\int d{\bf k} {m_\rho^*\over E_\rho ^*}
{1\over \exp (E_\rho ^*/T)-1}\nonumber\\
&+&{2g_S\over (2\pi )^3}\int d{\bf k} {m_\omega^*\over E_\omega ^*}
{1\over \exp (E_\omega ^*/T)-1}
+{6\sqrt 2 g_S\over (2\pi )^3}\int d{\bf k} {m_{a_1}^*\over E_{a_1}^*}
{1\over \exp (E_{a_1}^*/T)-1},
\end{eqnarray}
where we have used the constituent quark model relations for nucleon
and vector meson masses \cite{thomas94}, i.e.,
\begin{eqnarray}
m_N^*\approx m_N-g_S\langle S\rangle ,~m_\rho^*\approx
m_\rho-(2/3)g_S\langle S\rangle ,
~m_\omega^*\approx m_\omega -(2/3)g_S\langle S\rangle ,
\end{eqnarray}
the quark structure of $\eta$ meson ($\approx 0.58(u\bar u+d\bar d)-
0.57s\bar s$) \cite{holstein} in free space which leads to
\begin{eqnarray}
m_\eta^*\approx m_\eta -0.45g_S\langle S\rangle ,
\end{eqnarray}
and the Weinberg sum rule relation ($m_{a_1}\approx \sqrt 2 m_\rho$)
between the rho-meson and $a_1$ meson
masses \cite{wein67,kapu94}, i.e,
\begin{eqnarray}
m_{a_1}^*\approx m_{a_1}-(2\sqrt 2/3)g_S\langle S\rangle .
\end{eqnarray}
We have thus neglected in the present model the possible change of
$\eta$ (also $\eta^\prime$) properties at high temperature due
to the Debye-type screening of instanton effects, which 
has recently been studied in Ref. \cite{huang95,kapu95}.
 
The coupling constants $(g_V/m_V)$ and $(g_S/m_S)$
are determined by the nuclear matter ground state properties.
We use here the values corresponding to the original Walecka model 
\cite{qhd86} and solve the self-consistent condition to determine baryon and 
meson masses at a given baryon density and temperature. The results 
for the effective nucleon and rho meson masses are
shown in Fig. 1. It is seen that for a temperature
around 165 MeV and a baryon density of
about 0.4 fm$^{-3}$ that correspond to the initial conditions of our
fire-cylinder, the rho meson mass is reduced to about 270 MeV.
We would like to emphasize that the
effects of dropping vector meson masses on the dilepton spectrum
do not depend very much on the particular model one uses to describe the
in-medium vector meson properties, as long as all models give similar
density and temperature dependence for the vector meson masses. 
For example, Cassing {\it et al.} \cite{cass95} have obtained 
similar results as ours \cite{li95b} by assuming that 
the vector meson masses in dense matter are given by that 
from the QCD sum-rule calculations \cite{hatsu92}.
The choice of the present model is that it offers a consistent way to
treat the change of vector meson masses during heavy-ion collisions through
the change of the scalar potential. In this way, the total
energy of the system is always conserved. Note that 
with a stronger medium effect in the original Walecka model
we have been able to
lower the initial baryon density substantially  from that in 
Ref. \cite{li95b}, where the nonlinear Walecka model with scalar meson 
self-interactions has been used. 
The original high baryon density in Ref. \cite{li95b}
gave an unrealistically high density of hadrons.
 
The above formalism can be generalized to include strange hadrons, 
such as $K$, $\bar K$, and hyperons that 
contain also light quarks.  This is achieved by allowing the scalar and 
vector fields to couple to the light quarks in these hadrons.  In our model,
the masses of these particles are thus also reduced at finite temperature
and density. Since the phi meson does not consist of light quarks, its
mass is not modified in the present study. 
 
\section{The expanding fire-cylinder}
 
\subsection{Initial conditions}
 
Dileptons are produced in all stages of heavy-ion collisions at the
SPS/CERN energies. 
Since in this study we are chiefly concerned with dileptons 
with mass below 1 GeV, the initial Drell-Yan processes, that
contribute mainly to high-mass dileptons,  can be safely neglected.
Furthermore, contributions from the quark-gluon-plasma, if
it is formed in heavy-ion collisions at the SPS/CERN energies,
to low-mass dileptons as well as low-momentum photons, have been found
to be insignificant as compared to those from the hadronic phase
\cite{sinha95,gale95}. We thus start our transport model from
the expansion stage of heavy-ion collisions, which might be
in the chiral restoration phase as indicated by the
decrease of hadron masses. Very roughly speaking, this corresponds to the
mixed phase in descriptions which do not employ medium-dependent masses. 
The initial hadron abundance and their momentum distributions 
in this expanding fire-cylinder are constrained by requiring that the final 
proton and pion spectra and rapidity distributions reproduce 
the measured ones. Furthermore, results from 
microscopic simulations based on the string dynamics will also be used to
guide us in parametrizing the initial conditions
\cite{cassing,rqmd,sorge95}. We neglect in the present work the
isospin asymmetry of the colliding nuclei and assume that all charged
states of hadrons are equally populated.
 
\subsubsection{Baryon abundance}
 
Let us first discuss the initial baryon density in the fire-cylinder. For
a central S+Au collision, simple geometrical estimates would give about 90
nucleons from the target nucleus if all projectile nucleons 
are participants.  Of course, the number of
participant  nucleons from the target nucleus should be smaller
in heavy-ion collisions at the SPS energy of 200 GeV/nucleon
as a result of partial transparency and 
the not-exactly-zero impact parameter.  
From the experimental data on proton
rapidity distributions in the mid- to projectile-rapidity region,
which have been measured by the NA35 collaboration for S+Au collisions
\cite{na35a} and recently by the NA44 collaboration for S+Pb collisions
\cite{na44}, the total baryon number in
the fire-cylinder can actually be determined.
To describe reasonably the proton rapidity
distribution, we find that the baryon number in the
initial fire-cylinder is about 100 (32 from the projectile and
68 from the target). The center-of-mass rapidity of the fire-cylinder
is then 2.65, and this is consistent with the pion rapidity distribution
measured by the NA35 collaboration which shows a broad peak around
$y=2.6$-2.7 \cite{na35b}. We include all baryon
resonances with masses below 1720 MeV and also the low-lying
hyperons, i.e.,  $\Lambda$, $\Sigma$, $\Lambda (1405)$ and $\Sigma (1385)$.
In our model, the hyperons couple to the scalar and vector fields 
with 2/3 of the strength of non-strange baryons, since they contain 
2 light quarks instead of 3 in non-strange baryons.
Their initial abundances
are determined by assuming strangeness saturation as
in Refs. \cite{pbm95}. Thus knowing the initial baryon
density, these abundance can be uniquely determined.
 
Initially, all baryons are distributed
in a fire-cylinder whose cross section is taken to be about 40 fm$^{2}$,
similar to the geometrical cross section of the projectile nucleus. If we
further assume that the initial baryon density is about 2.5$\rho _0$
($\rho _0 = 0.16$ fm$^{-3}$), then the initial
longitudinal length $2z_L$ of the fire-cylinder is found to be 6.2 fm.
The initial volume of the fire-cylinder is thus about 250 fm$^{-3}$.  
 
\subsubsection{Meson abundance}
 
For mesons, we include $\pi$, $\eta$, $\eta^\prime$, $\rho$, $\omega$, 
$\phi$, and $a_1$, as well as $K$ and $K^*(892)$. 
In the earlier stage of heavy-ion collisions preceding our initial
fire-cylinder, mesons are expected to be copiously produced
either from string fragmentation as in the RQMD \cite{rqmd,sorge95} or
from the hadronization of a quark-gluon plasma as in \cite{bert95}.
They can also be produced from processes like nucleon-antinucleon 
annihilation in a chirally restored phase \cite{brown93}.  
The initial abundance of (non-strange) mesons (mainly $\pi$, $\rho$, 
$\omega$ and $a_1$ that feed appreciably to the final pion multiplicity)
in the fire-cylinder is determined by 
requiring that the final pion number agrees with the measured value 
and will be specified later. 
 
For the meson properties in hot and dense hadronic matter, we will use
two models. In the first model, all mesons are taken to have free masses,
and their abundances are determined by assuming that they are in
thermal and chemical equilibrium, i.e., the 
$\pi$, $\rho$, $\omega$, and $a_1$ chemical potentials are related by
$\mu _\rho =2\mu _\pi$ and $\mu _{a_1}=\mu _\omega =3\mu _\pi$.
At a given temperature and pion chemical potential, we can determine 
the meson densities, and from which the meson multiplicities are calculated
using the initial volume of the fire-cylinder as determined above.
Normally, one would expect that $\mu_\pi =0$ so that pions 
are also in chemical equilibrium with baryons, which is indeed the case
at the freeze-out as shown in Ref. \cite{pbm95}. However,
we find that in order to reproduce the observed pion rapidity 
distribution and spectrum requires an initial pion chemical potential 
of about 130 MeV at an initial temperature of 165 MeV. This temperature
is determined in our model by fitting the slopes of the proton and 
pion transverse momentum spectra after the full dynamical evolution 
of the system. Thus, in the initial stage of our simulation, mesons
are out of chemical equilibrium with baryons, and they approach chemical 
equilibrium as the system expands. Microscopic simulations
based on the RQMD show that in the central region of heavy-ion
collisions the pion density after initial string fragmentations
is about 0.4 fm$^{-3}$, and the pion temperature is about 170 MeV 
\cite{li96}.
This density is nearly double the equilibrium one and 
indicates an appreciable pion chemical potential for those pions which
come from string fragmentation.
The initial $\pi$, $\rho$, $\omega$ and $a_1$ meson
numbers are then determined to be 119, 56, 43, and 10, respectively.
 
Because of finite decay width, the vector meson mass 
is given the following distribution
\begin{equation}\label{mass}
P(M)\sim\frac{(m\Gamma)^2}{(M^2-m^2)^2+(m\Gamma)^2},
\end{equation}
with a proper normalization. In the above, $m$ denotes
the centroid mass of the meson while $\Gamma$ is its width evaluated
at a mass $M$. 
 
For rho and phi mesons, their decay widths
$\rho\rightarrow \pi\pi$ 
and $\phi\rightarrow K\bar K$ are given, respectively, by \cite{li95a}
\begin{eqnarray}\label{rpp}
\Gamma _{\rho\rightarrow \pi\pi} (M)=
{g^2_{\rho\pi\pi}\over 4\pi} {\big(M^2-4m_\pi^2\big)^{3/2}\over 12 M^2},
\end{eqnarray}
and
\begin{eqnarray}
\Gamma _{\phi\rightarrow K\bar K} (M)=
{g^2_{\phi K \bar K}\over 4\pi} {\big(M^2-4m_K^2\big)^{3/2}\over 6 M^2}.
\end{eqnarray}
where $g_{\rho\pi\pi}^2/4\pi \approx 2.9$ and $g_{\phi K\bar K}^2/4\pi
\approx 1.7$ are determined from the measured widths at $m_\rho\approx
770$ MeV and $m_\phi\approx 1020$ MeV, respectively.
 
The decay width for $K^*\rightarrow K\pi$ has been derived in Ref. \cite{ko94}
to be
\begin{eqnarray}
\Gamma _{K^*\rightarrow K\pi}(M)={g_{K^*K\pi}^2\over 4\pi}
{2k^3\over M^2},
\end{eqnarray}
where $k$ is the momentum of the pion in the center-of-mass frame 
of $K^*$, and $g^2_{K^*K\pi} / 4\pi\approx$ 0.86 is determined from the
measured $K^*$ decay width at $m_{K^*}\approx 892$ MeV. 
For the decay width of $a_1\rightarrow
\pi\rho$, we use the result of Ref. \cite{xiong92}, i.e.,
\begin{eqnarray}
\Gamma _{a_1\rightarrow \pi\rho}={G_{a_1\pi\rho}^2 k\over
24 \pi m_{a_1}^2}\big[2(p_\pi\cdot p_\rho )^2+m_\rho ^2(m_\pi^2+k^2)\big],
\end{eqnarray}
where $k$ is the pion momentum in the rest frame of $a_1$, and 
$G_{a_1\pi\rho}\approx 14.8$ GeV$^{-1}$ is determined from the $a_1$ decay
width in free space using its centroid mass.
There is, unfortunately, no simple expression for the decay width of
$\omega\rightarrow \pi^+\pi^-\pi^0$ \cite{krs84}. 
We use in this work the approximation
that this width is directly proportional to the mass of the omega meson,
which becomes exact in the chiral limit of $m_\pi\rightarrow 0$.
 
Since we will consider the $\eta$ and $\eta^\prime$
Dalitz decay contributions to the dilepton
yield, their initial multiplicities need to be determined.
As for pions, the best way is to fit
the experimental data after dynamical evolution of the system.
There are published data from the WA80 collaboration for minimum-biased
events and preliminary data for central events in 
S+Au collisions at 200 GeV/nucleon \cite{wa80a,wa80b,wa80c}. 
Both are given in terms
of $\eta /\pi^0$ ratio as a function of the transverse momentum. 
We thus determine the initial $\eta$ and $\eta^\prime$
number mainly by fitting the final $\eta /\pi^0$ 
ratio to the measured one.
In the chemical equilibrium scenario for the meson chemical composition,
both the $\eta$ and $\eta^\prime$ chemical potentials 
are assumed to be about three times that of the pion.
The large eta chemical potential is also consistent with that of 
RQMD calculations \cite{li96}.  This gives us about 37 $\eta$ mesons
and 5 $\eta^\prime$ mesons in the initial fire-cylinder.
The final $\eta$ meson number at the freeze-out is about
32 (see Fig. 6 below).
This includes the primary $\eta$ mesons that have not been absorbed
by nucleons or converted into pions through $\eta\eta\rightarrow \pi\pi$,
$\eta$ mesons from $N^*(1535)$ which decay into $\eta N$ with a branching
ratio of about 0.4 \cite{pada},
$\eta$ mesons from $\eta^\prime$ which decays into
$\pi\pi\eta$ with a branching ratio of about 2/3 \cite{pada},
and $\eta$ mesons produced from $\pi\pi\rightarrow \eta\eta$.
We find that the experimental $\eta/\pi^0$ ratio is reasonably 
reproduced in this case (see Fig. 10 below).
As we will show later, in CERES experiments 
the $\eta^\prime$ contribution to dileptons
is unimportant but that from $\eta$ dominates the 
dilepton spectrum at invariant mass between $150<M<350$ MeV. 
 
In the second model for the meson properties, 
we take into account the effects of medium modifications according 
to that predicted
by the extended Walecka model as discussed in the previous section.
At an initial temperature of 165 MeV and an initial baryon density
of 2.5$\rho_0$, we find that the rho and omega meson masses
reduce to about
270 MeV, while the mass of the $a_1$ meson reduces to about 550 MeV.
As a result, their abundance in the initial fire-cylinder increases,
and we need essentially a zero pion chemical potential 
in order to reproduce the experimental pion rapidity distribution 
and transverse momentum spectrum. 
We expect a  similar trend in RQMD as a part of the energy is now 
converted to the field energy, so less energy is available for 
pion production, which is, however, compensated by an increase of
rho, omega, and $a_1$ production due to reduced masses.
The initial $\pi$, $\rho$, $\omega$, 
and $a_1$ meson numbers are then determined to be 44, 92, 33, and 26, 
respectively. The initial low-mass vector and axial-vector 
mesons thus act as a reservoir for the observed large pion yield. 
As in the free mass case, the mass distributions for these mesons
are given by Eq. (\ref{mass}) with the centroid mass
in the equation replaced by the 
in-medium one.
 
In the extended Walecka model, the masses 
of $\eta$ and $\eta^\prime$ (using $\eta^\prime\approx 0.40(u\bar u+d\bar d)
+0.82s\bar s$, we have $m_{\eta ^\prime}^* \approx m_{\eta ^\prime}
-0.2\langle S \rangle $ \cite{holstein}) in the initial stage
are reduced to about 210 and 810 MeV, respectively.
Because of the large strange quark content in $\eta^\prime$,
the reduction of its mass is less significant than
that of the $\eta$ meson mass as also noted in Ref. \cite{huang95}.
When the effective masses of $\eta$ and $\eta^\prime$ 
are reduced in hot dense matter, either by the screening of instanton effects
as in Refs. \cite{huang95,kapu95} or by the attractive scalar field as
in our model, their initial chemical potentials can be significantly reduced.
We are, however, unable to reduce them to zero  
as for the pion chemical potential due to the smaller contribution to 
$\eta$ production from higher meson resonances than to pion production.
The inclusion of $U_A(1)$ restoration \cite{huang95,kapu95} and $a_0(980)$,
that decays dominantly into $\eta\pi$, will modify this. 
In addition, since the eta meson mass is reduced in hot dense matter, the
process of eta production in pion-pion interaction $\pi\pi\rightarrow
\eta\eta$ is favored over the reverse one. We find that in order to 
describe well the experimental $\eta /\pi^0$ ratio (see Fig. 10 below), 
we need initially an eta chemical potential of about 130 MeV.
The initial $\eta$ and $\eta^\prime$ numbers are then found to be
26 and 3, respectively, and the final $\eta$ number at the freeze-out is
about 32.
 
\subsubsection{Momentum distributions}
 
The initial transverse momenta of all hadrons are assumed to 
be given by a thermal distribution, with a proper slope parameter
that is loosely referred to as `temperature' in the transverse direction.   
We find that an initial temperature of about 165 MeV is needed 
to reproduce the observed slopes of the transverse mass spectra 
for both protons and pions. This is somewhat lower than that 
used in Ref. \cite{li95b} as a result of the inclusion of additional degrees 
of freedom.  Introducing more degrees of freedom should 
lower it further. The longitudinal momentum distribution is 
determined by imposing a rapidity field as in the hydrodynamical 
model \cite{bolz92,venu94}. Specifically, we assume 
that the rapidity of a particle in the fire-cylinder frame
is correlated to its longitudinal position via the following 
gaussian function, i.e.,
\begin{eqnarray}\label{corr}
f(y,z)={1\over \sqrt {2\pi} \sigma}\exp\big(-(y-y_Lz/z_L)^2/2\sigma ^2\big).
\end{eqnarray}
 
For mesons, RQMD simulations show that their initial rapidity 
distributions depend on the mass, with pions (the lightest) having 
the broadest rapidity distribution \cite{li96}.  For simplicity, 
we will use the same $y_L$ and $\sigma$ in Eq. (\ref{corr}) for all mesons. 
We find that with $y_L=0.6$ we have a good description
of the experimental pion rapidity distribution, particularly in the
mid-rapidity region (see Fig. 8 below), if the variance $\sigma$ is chosen 
to be 0.8 (also for baryons) so that there is substantial
dispersion in the $y$-$z$ correlation.  Thus, mesons are initially 
situated mostly around the mid-rapidity but 
becomes broader at the freeze-out due to interactions, leading 
to good agreements with the experimental data in the mid-rapidity region. 
 
We note that the initial rapidity distributions in our model is consistent
with a momentum distribution that is in
local thermal equilibrium at a temperature of about 165 MeV.
The introduction of a rapidity field, 
however, breaks the equilibrium between the longitudinal and transverse 
momentum distributions. This is again consistent with that from the 
RQMD \cite{li96}.
 
Baryons, on the other hand, show two components in the initial
rapidity distribution \cite{shuryak}. 
The lower component, with $y_L=1.0$,
is similar to the mesons, while the upper component, due to limiting
fragmentation, has a larger $y_L=2.0$.
The initial baryon rapidity distribution is thus broader
than that of mesons. This is necessary
in order to fit the observed broad proton rapidity distribution,
as the proton rapidity distribution is hardly
changed by their interactions with other particles as a result of their
large mass. In Fig. 2, we show the initial rapidity and transverse 
momentum distributions of protons and negative pions for the case of 
free masses.
 
\subsection{The relativistic transport model}
 
The expansion of the fire-cylinder is treated by the 
relativistic transport model \cite{ko87,li94}. In this model, 
hadrons (except for pions whose medium modifications
are neglected in present work) are propagated in their
mean-field potential. For a nucleon, this is
\begin{eqnarray}
{d{\bf r}\over dt}={{\bf k}\over E_N^*}, ~~{d{\bf k}\over dt}=
-\nabla _r(E_N^*+(g_V/m_V)^2\rho _B),
\end{eqnarray}
Similar equations are used for baryon resonances with their mean-field
potentials taken to be the same as the nucleon one.  Furthermore, 
particles also undergo stochastic two-body collisions.
For baryon-baryon interactions, we include both
elastic and inelastic scattering for nucleons, $\Delta (1232)$,
$N(1440)$ and $N(1535)$. Their cross sections are either taken from
Refs. \cite{wolf93,va82} or obtained using the standard detailed-balance
procedure \cite{db91}. Lacking empirical information,
we consider only elastic scattering for interactions involving higher
baryon resonances and hyperons. This is a reasonable approximation as nucleon,
$\Delta (1232)$, $N(1440)$ and $N(1535)$ account for about 80\% of all baryons.
Moreover, in our thermalized fire-cylinder, the energy of a colliding baryon
pair is not very large, so higher resonance production from baryon-baryon
interaction is unimportant. Moreover, their production cross sections 
in baryon-baryon interactions are about an 
order of magnitude smaller than those in meson-baryon interactions, 
which are included in our model through resonance
formation and decay. For example, the interaction of a pion with
a nucleon proceeds through
the formation of $\Delta (1232)$, $N(1440)$, $\cdots$, $N(1720)$.
The formation cross sections are taken to be of the relativistic Breit-Wigner
form, e.g., the isospin-averaged cross section 
$\sigma _{\pi N\rightarrow \Delta } (\sqrt s)$ is given by \cite{li94},
\begin{eqnarray}
\sigma _{\pi N\rightarrow \Delta}(\sqrt s)
={16\pi\over 3k^2} {(m_\Delta \Gamma _\Delta)^2
\over (s-m_\Delta ^2)^2+(m_\Delta \Gamma _\Delta )^2},
\end{eqnarray}
where $k$ is the pion momentum in the center-of-mass frame of the $\Delta$
particle.
 
The meson-meson interactions are either formulated by the resonance formation
and decay when the intermediate meson is explicitly included in our model
or treated as a direct elastic scattering with a constant cross section
estimated from various theoretical models.
 
For a pair of pions with a total invariant mass $M$, a rho meson
of this mass is formed with an isospin-averaged
cross section given by the Breit-Wigner form \cite{li95a},
\begin{eqnarray}\label{ppr}
\sigma _{\pi \pi \rightarrow \rho } (M) = {8\pi\over k^2}
{(m_\rho \Gamma _\rho )^2\over (M^2-m_\rho ^2)^2+(m_\rho \Gamma _\rho )^2}
\big({M\over m_\rho}\big)^2,
\end{eqnarray}
where $k$ is the pion momentum in the center-of-mass frame of 
rho meson, and the multiplying factor $(M/m_\rho )^2$ will be clarified 
later. Similarly, for
$\pi\rho \rightarrow a_1$ and $\pi K\rightarrow K^*$, 
the isospin-averaged cross sections \cite{xiong92,ko81} are, respectively,
\begin{eqnarray}
\sigma _{\pi \rho \rightarrow a_1 } (M) = {4\pi\over 3k^2}
{(m_{a_1} \Gamma _{a_1} )^2\over (M^2-m_{a_1} ^2)^2+(m_{a_1} \Gamma _{a_1})^2},
\end{eqnarray}
and
\begin{eqnarray}\label{pkk}
\sigma _{\pi K \rightarrow K^* } (M) = {4\pi\over k^2}
{(m_{K^*} \Gamma _{K^*} )^2\over (M^2-m_{K^*} ^2)^2+(m_{K^*} \Gamma _{K^*})^2}.
\end{eqnarray}
 
Interactions of pions with other mesons are treated through direct
elastic scattering. The energy-averaged 
cross section for $\pi\eta \rightarrow \pi\eta$, which is dominated by
$a_0(980)$ formation and decay, 
is estimated from the amplitudes given in Ref. \cite{huang95} and 
is about 20 mb, and the same is assumed for $\pi\eta^\prime\rightarrow
\pi\eta^\prime$.
The energy-averaged cross section for $\pi\omega\rightarrow \pi\omega$
is estimated from the omega collisional broadening width due to the pion-omega
interaction in Ref. \cite{hag95} and is about 15 mb, and the same is assumed
for $\pi K^*\rightarrow \pi K^*$ and $\pi a_1\rightarrow \pi a_1$.
Finally, we have also included the following inelastic processes for
pion-pion collisions, $\pi\pi\rightarrow \eta\eta$ and $\pi\pi\rightarrow 
K\bar K$. The cross section for the former is again obtained from Ref.
\cite{huang95} which is about 10 mb, while the cross section
for the latter has been calculated in Ref. \cite{ko90} based on the
one-boson exchange model and is about 1-3 mb depending on 
the incident energy.
 
The energy-averaged cross section for $\eta\eta\rightarrow \eta\eta$ is
estimated using amplitudes given in Ref. \cite{huang95} and is about
12 mb. Interactions of an eta with other mesons are assumed to have 
similar cross sections. 
The inelastic cross section for $\eta\eta\rightarrow \pi\pi$ is obtained
from $\pi\pi\rightarrow \eta\eta$ using detailed balance.
 
The collision of a kaon with a pion is treated through the 
$K^*(892)$ formation and
decay as discussed earlier. The kaon-antikaon collision mainly 
proceeds through
the formation and decay of the phi meson, and the isospin-averaged
cross section is given by \cite{li95a}
\begin{eqnarray}\label{kap}
\sigma _{K\bar K \rightarrow \phi } = {3\pi\over k^2}
{(m_{\phi} \Gamma _{\phi} )^2\over (M^2-m_{\phi} ^2)^2+(m_{\phi} 
\Gamma _{\phi})^2}\big({M\over m_\phi}\big)^2,
\end{eqnarray}
The cross section for the direct elastic scattering of a kaon and a 
rho meson
is estimated to be about 30 mb from the rho meson collisional broadening
width in Ref. \cite{hag95}. This cross section is relatively large due
to the presence of a broad $K^*(1270)$ resonance right above the $K\rho$
threshold. All other meson-meson collisions are treated as direct
elastic processes with a constant cross section of 20 mb, which is
similar to that used in Ref. \cite{bert95}.
 
The above discussions apply to both free and 
medium-dependent masses. In the latter, the centroid masses 
(e.g., $m_\rho$ in Eq. (\ref{ppr})) in these equations are replaced by the 
in-medium ones, while the decay widths (e.g. Eq. (\ref{rpp})) are evaluated
with the physical mass $M$ of mesons.
 
\subsection{Time evolution of the fire-cylinder}
 
As an illustration, we show in Fig. 3
the baryon (left panel) and meson (right panel)
distributions in the x-z plane at several time steps.
Initially all particles are located in a compressed cylinder
whose transverse length is slightly larger than the longitudinal 
one. Then the system expands longitudinally as well as transversely
with corresponding flow velocities $\beta _z$ and $\beta _t$, 
\begin{equation}
\beta_z(z)=\frac{1}{N_z}\sum_{i=1}^{N_z}\frac{p_{iz}}{E_i},\qquad
\beta_t(r_t)=\frac{1}{N_t}\sum_{i=1}^{N_t}\frac{p_{it}}{E_i},
\end{equation}
where the summation in the first expression is over all $N_z$ particles 
with longitudinal coordinates $(z-0.5,z+0.5)$ fm, and in the second 
expression it is over all $N_t$ particles with
transverse coordinates $(r_t-0.5,r_t+0.5)$ fm ($r_t=\sqrt {x^2+y^2}$).
The results are shown in Fig. 4 at the freeze-out time (t=20 fm/c).
It is seen that the longitudinal flow velocity is essentially 
linear in z. For small $r_t$, the transverse
flow velocity is also approximately linear in the transverse 
distance.
 
The time evolution of the baryon abundance is shown in Fig. 5. During 
initial expansion, we see an increase of the abundance of 
higher baryon resonances. This is mainly due to the fact that 
pions are initially 
out of chemical equilibrium with baryons, and higher baryon
resonance formation is thus 
favored over the reverse process. As the
system expands, baryon resonances eventually decay into pions and nucleons.
We do not see much difference between the results using free
meson masses (the left panel) and in-medium  meson masses (the right
panel) except that the increase of higher baryon resonances
is less significant in the latter case. This is reasonable as in this
case the initial pion density is smaller.
A similar plot is shown in Fig. 6 for the time evolution of the
meson abundance.
 
The time evolution of the central baryon density of the fire-cylinder 
is shown in Fig. 7, and it is seen that the
system expands slower when in-medium meson masses are used.
In this case, some of the energy goes into the field energy 
$(m_S^2/2)\langle S \rangle ^2$. 
Since this energy is independent of baryon number, the 
associated pressure is negative and thus reduces the kinetic pressure.
This is not a large effect at SPS energies but will
be very important at AGS energies, where the fraction of energy in
the field is appreciable. We plan to work this out quantitatively
for AGS energies, in the hope that dileptons can be measured at some time
in the future in this energy domain.
The freeze-out temperature of the system, extracted from the pion 
transverse mass spectrum by removing the effect from transverse flow,
is about 130 MeV. The apparent slope parameters
of the pion and proton transverse mass spectra are, however, larger
as a result of transverse expansion.
 
The final proton and pion transverse momentum spectra
are compared with the experimental data in Fig. 8. The results 
using free meson masses are shown by the dotted histograms, 
while those using in-medium masses are shown by the solid histograms.
The pion transverse momentum spectra in both cases
are seen to be in good agreement with the measured data from the 
CERES \cite{ceres94} and the WA80 \cite{wa80a} collaboration
which have been presented with arbitrary normalizations.
The inverse slope parameters of the pion and proton
transverse mass spectra have been extracted from our calculation;
they are, respectively, 150 and 220 MeV and are in good agreement with
those extracted by the NA44 collaboration for central S+Pb
collisions at 200 GeV/nucleon \cite{na44}.
 
In Fig. 9, we compare the pion and proton rapidity distributions
with the experimental data. In our calculation we have not removed
the protons from the hyperon decay as in the NA44 collaboration \cite{na44}.
We have thus added the hyperons back to the proton rapidity
distribution given in Ref. \cite{na35a} by the NA35 collaboration in
order to make a meaningful comparison. In the mid-rapidity region
covered by the CERES collaboration, our pion rapidity distribution 
is in good agreement with the data from the NA35 collaboration \cite{na35b}.
Since we have not attempted to describe the physics around the target
rapidity region, which does not affect our study of dilepton production
in the mid- to projectile-rapidity region, our pion rapidity distribution in the
small $y$ region should be compared with the open squares which are
obtained by reflecting the experimental data at large $y$
with respect to $y_{cm}\approx
2.65$. In the mid-rapidity region our pion rapidity distribution also
agrees with the preliminary data from the NA44 collaboration for
central S+Pb collisions \cite{na44}.
 
In Fig.10, we compare the $\eta /\pi^0$ ratio obtained in our model
with the experimental data from the WA80 collaboration \cite{wa80b,wa80c}.
For the minimum-biased events (solid circles) the data have been 
officially published \cite{wa80c}, while for  central collisions
(solid squares) the data are very preliminary \cite{wa80b}.
Our $\eta /\pi^0$ ratio is larger than the minimum-biased data
but slightly smaller than the central collision data, and this is
especially so in the low transverse momentum region. We believe that
this corresponds reasonably to the $\eta /\pi^0$ ratio in the CERES
experiments.  With an average charged-particle multiplicity
of $dN_{ch}/d\eta \approx$ 125 in the rapidity range $2.1<\eta <2.65$,
the CERES experiments select a smaller centrality than
the WA80 experiments which have an average charged-particle multiplicity
of more than 160 in the same rapidity region \cite{wa80b,wa80d}.
This gives us the confidence that
our results for the dilepton production from the $\eta$ and $\eta^\prime$
Dalitz decay are quantitatively correct, since their contribution to the 
dilepton yield is essentially determined by
their freeze-out multiplicities as a result of their much longer 
lifetime as compared to the lifetime of the fire-cylinder or 
the nucleus-nucleus
interaction time (see discussions in next section and Fig. 18 below).
 
By fitting and extrapolating the measured $\eta$ and $\pi ^0$ yields
to the full
phase space, an $\eta /\pi^0$ number ratio of 0.147$\pm$0.017(stat.)$\pm$
0.015(syst.) has been obtained for the minimum-biased S+Au collisions by
the WA80 collaboration \cite{wa80c}. 
This is to be compared with a ratio of about 0.08 in
the proton-proton interaction at similar center-of-mass energies \cite{lebc}.
On the other hand, it has been estimated in Ref. \cite{wa80b}
that for the 9\%  most central S+Au collisions this ratio could be
twice that in minimum-biased collisions, implying an $\eta/\pi^0$ number
ratio of about 0.29 in the 9\%  most central collisions. 
The $\eta /\pi^0$ number ratio in
the CERES experiments should then be somewhere between these two limits,
and our $\eta /\pi^0$ number ratio of about 0.2 is thus quite reasonable.
 
\section{dilepton production}
 
The main contributions to dileptons with mass below 1 GeV are the
Dalitz decay of $\pi^0$, $\eta$ and $\omega$, the direct leptonic decay of
vector mesons such as $\rho^0$ and $\omega$, and the pion-pion annihilation
which proceeds through the $\rho^0$ meson by the vector dominance. In addition,
we will also evaluate dilepton production from $\eta^\prime$ and $a_1$
decay as well as from $\phi$ decay and kaon-antikaon annihilation.
 
\subsection{Dalitz decay}
 
The Dalitz decays of $\pi^0$, $\eta$, and $\omega$ contribute significantly
to dileptons with mass below 2$m_\pi$. The differential width
of the Dalitz decay can be related to its radiative decay width \cite{land85}.
For example, for $\pi^0\rightarrow \gamma e^+e^-$, we have \cite{land85}
\begin{eqnarray}
{d\Gamma (\pi^0\rightarrow \gamma e^+e^-)\over dM}&
=&{4\alpha\over 3\pi}{\Gamma (\pi^0\rightarrow 2\gamma)\over M}
\Big(1-{4m_l^2\over M^2}\Big)^{1/2}\nonumber\\
&\times& \Big(1+{2m_l^2\over M^2}\Big)
\Big(1-{M^2\over m_{\pi^0}^2}\Big)^3|F_{\pi^0}(M^2)|^2,
\end{eqnarray}
where $M$ is the mass of the produced dilepton, $\alpha$ is the fine structure
constant, $\Gamma (\pi^0\rightarrow 2\gamma)\approx$ 7.65 eV
is the neutral pion radiative decay width \cite{pada}, and $m_l$ is
the mass of the lepton. In the case of dielectron production, 
$m_l=m_e\approx 0.51$ MeV can be neglected. 
The electromagnetic form factor is parameterized as
\begin{eqnarray}
F_{\pi^0}(M^2)=1+b_{\pi^0}M^2,
\end{eqnarray}
with $b_{\pi^0}=5.5$ GeV$^{-2}$ as determined empirically \cite{land85}.
For this process the form factor effects are small as 
the dilepton invariant mass is always small ($M<m_{\pi ^0}$).
 
Similarly, for $\eta\rightarrow \gamma e^+e^-$, we have
\begin{eqnarray}\label{egee}
{d\Gamma (\eta\rightarrow \gamma e^+e^-)\over dM}&
=&{4\alpha\over 3\pi}{\Gamma (\eta\rightarrow 2\gamma)\over M}
\Big(1-{4m_l^2\over M^2}\Big)^{1/2}\nonumber\\
&\times& \Big(1+{2m_l^2\over M^2}\Big)
\Big(1-{M^2\over m_{\eta}^2}\Big)^3|F_{\eta}(M^2)|^2,
\end{eqnarray}
where $\Gamma (\eta\rightarrow 2\gamma)\approx 0.463$ KeV is the
radiative decay width of an eta meson \cite{pada}. The electromagnetic form
factor of an eta meson is expressed in the pole approximation as
\begin{eqnarray}
F_{\eta}(M^2)=\Big(1-{M^2\over \Lambda_{\eta}^2}\Big)^{-1},
\end{eqnarray}
with the cut-off parameter $\Lambda_{\eta}\approx 0.72$ GeV as
determined experimentally from $\eta\rightarrow \gamma \mu ^+\mu ^-$
\cite{land85}. This value is close to that predicted by the 
vector dominance model and used in Ref. \cite{wolf93}. 
In this case the form factor effect is important for dileptons 
with mass close to $m_\eta$.
 
For $\omega\rightarrow \pi^0e^+e^-$, the differential decay width is
given by
\begin{eqnarray}\label{opee}
{d\Gamma (\omega\rightarrow \pi^0 e^+e^-)\over dM}&
=&{2\alpha\over 3\pi}{\Gamma (\omega\rightarrow \pi^0\gamma)\over M}
\Big(1-{4m_l^2\over M^2}\Big)^{1/2}\Big(1+{2m_l^2\over M^2}\Big)
\nonumber\\
&\times&\Big[\big(1+{M^2\over m_\omega^2-m_{\pi^0}^2}\big)^2
-\big({2m_\omega M\over m_\omega^2-m_\pi^2}\big)^2\Big]
|F_{\omega\pi^0}(M^2)|^2,
\end{eqnarray}
where $\Gamma (\omega\rightarrow \pi^0\gamma)=0.717$ MeV is the omega
meson radiative decay width \cite{pada}. The electromagnetic form
factor is similar to Eq. (\ref{egee}) with $\Lambda _\eta$ replaced by
$\Lambda_\omega$.
In the vector dominance model, the cut-off parameter $\Lambda _{\omega}$
would have a value of about 0.77 GeV, the rho-meson mass. However, 
from the omega muonic decay \cite{land85} its value is found
to be around 0.65 GeV, which will be used in the present work.
The use of the vector dominance
model prediction would underestimate the dileptons with mass close
to $m_\omega-m_{\pi^0}$.
 
For $\eta^\prime\rightarrow \gamma e^+e^-$, we have a formula similar to Eq.
(\ref{egee}). 
However, in this case the vector meson pole ($m_\rho$) occurs in
the physical region allowed for the dilepton spectrum
($M<m_{\eta^\prime}$), and the simple pole approximation
is not valid. Instead we use the original vector-dominance form factor
\cite{li95a},
which has been shown to reproduce the experimental data reasonably well
for $\eta^\prime \rightarrow \gamma\mu^+\mu^-$ \cite{land85}.
The treatment of the $a_1$ Dalitz decay is slightly different. As we have in
our dynamical model explicitly the processes $a_1\leftrightarrow\pi\rho$
and $\rho\rightarrow e^+e^-$, the part of $a_1$
contribution to dileptons which proceeds through the physical $\rho$ meson
as a two-step process has already been included. Thus, 
in evaluating the $a_1$ Dalitz decay ($a_1\rightarrow\pi e^+e^-$)
contribution we need not introduce the vector-dominance model form factor.
Otherwise, there would be double counting.
 
\subsection{Vector meson decay into dileptons}
 
The direct leptonic decay of vector mesons is an important source of 
dileptons. For the dilepton invariant mass region of interest to this work,
we consider mainly the leptonic decay of $\rho ^0$,  $\omega$ and
$\phi$ mesons.
The decay width for $\rho ^0\rightarrow e^+e^-$ is given by
\begin{eqnarray}\label{ree}
\Gamma _{\rho^0\rightarrow e^+e^-}(M)={g_{\rho\gamma}^2e^2\over M^4}{M\over 3}
={\alpha^2\over \big(g_{\rho\pi\pi}^2/4\pi \big)} {m^4_\rho \over 3M^3},
\end{eqnarray}
where $M^4$ in the denominator arises from the virtual photon propagator,
and $M$ in the numerator comes from the phase space integration. In obtaining
the second expression,
we have used the vector dominance
relation $g_{\rho\gamma}=em_\rho^2/g_{\rho\pi\pi}$ \cite{bhaduri88}. 
In the calculation, we use, however, $\Gamma _{\rho ^0\rightarrow
e^+e^-}(M) =8.814\times 10^{-6}m_\rho^4/M^3$, with the coefficient
determined from the measured width at $M=m_\rho$. Similarly, we have  
$\Gamma_{\omega\rightarrow e^+e^-} (M) =0.767 \times 10^{-6}m_\omega^4/M^3$
and $\Gamma_{\phi\rightarrow e^+e^-} (M) =1.344 \times 10^{-6}m_\phi^4/M^3$
for the omega and phi meson dieletron decay widths, respectively.
 
Combining Eqs. (\ref{rpp}), (\ref{ppr}), and (\ref{ree}), we obtain
\begin{eqnarray}\label{ppee}
\sigma _{\pi^+\pi^-\rightarrow \rho^0 \rightarrow e^+e^-} (M)
=\sigma _{\pi ^+\pi ^-\rightarrow \rho ^0} \cdot \Gamma _{\rho^0\rightarrow
e^+e^-}/\Gamma _{\rho}
={8\pi \alpha^2 k\over 3M^3} |F_\pi (M)|^2,
\end{eqnarray}
with the pion electromagnetic form factor given by
\begin{eqnarray}\label{form}
|F_\pi (M)|^2 = {m_\rho ^4\over (M^2-m_\rho ^2)^2+(m_\rho \Gamma _\rho )^2}.
\end{eqnarray}
This is the dilepton production cross section from pion-pion annihilation
commonly used in the form factor approach
\cite{gale87,li95a,xiong90,wolf93}. 
Thus, the dilepton yield from the pion-pion annihilation obtained in
the dynamical approach with explicit rho meson formation
agrees with that obtained in the form factor approach if
medium effects on the intermediate rho meson are neglected \cite{li95a}.
This explains the multiplying factor $\big(M/m_\rho \big)^2$
in Eq. (\ref{ppr}) (and a similar factor for kaon-antikaon
annihilation, Eq. (\ref{kap})). In Refs. \cite{li95a,li95b} we have 
used the rho meson dilepton decay width $\Gamma _{\rho^0\rightarrow e^+e^-} 
(M)= 1/3 \alpha ^2 M (4\pi /g_{\rho\pi\pi}^2)$. In order for
the relation Eq. (\ref{ppee}) to hold, we thus need a somewhat complicated
multiplying factor for the simple Breit-Wigner cross section
as shown in Eqs. (\ref{pkk}) and (\ref{form}) of Ref. \cite{li95a}. 
Neglecting the medium modification of intermediate vector mesons,
the present way of separating $\sigma _{\pi^+\pi^-\rightarrow \rho^0}$ and 
$\Gamma _{\rho^0\rightarrow e^+e^-}$ and the way used in Refs.
\cite{li95a,li95b} both give the same results as the form factor approach
for the pion-pion (and similarly kaon-antikaon) contribution.  
For contributions from the primary rho and omega mesons,
different forms of $\Gamma _{\rho^0, \omega \rightarrow e^+e^-}$ as used
here and in Ref. \cite{li95b}
do give rise to differences in dilepton yields with masses far away from 
the rho and omega centroid masses. Since low-mass
dileptons are dominated by pion-pion annihilation and at high masses
kaon-antikaon annihilation becomes significant, we find that
the results of Ref. \cite{li95b} are only slightly affected 
when the leptonic decay width with an explicit photon propagator is used.
 
When medium modifications of vector meson masses
are included, the rho meson mass
$m_\rho$ in Eqs. (\ref{rpp}) and (\ref{ree}) is replaced by 
the in-medium mass $m_\rho^*$.  In a static hadronic matter, one obtains
from the relation $\sigma _{\pi^+\pi^-\rightarrow \rho^0 \rightarrow 
e^+e^-} (M)=\sigma _{\pi ^+\pi ^-\rightarrow \rho ^0} 
\cdot \Gamma _{\rho^0\rightarrow e^+e^-}/\Gamma _{\rho}$ 
again Eqs. (\ref{ppee}) and (\ref{form}) with $m_\rho$ replaced by $m_\rho^*$. 
We note that in obtaining this result the in-medium vector 
dominance relation,
$g_{\rho\gamma}^*=em_\rho^{*2}/g_{\rho\pi\pi}$, is assumed. With reduced
rho meson mass in medium, the vector dominance is thus weakened.
This effect has
also been shown in studies based on the hidden gauge theory
including loop corrections at finite temperature \cite{songa}.
 
Another appreciable medium effect is the collisional broadening of vector
meson widths in medium. Haglin \cite{hag95} has shown that,
at a temperature of
150 MeV, the collisional widths are about 50 MeV and 30 MeV
for the rho and omega mesons, respectively.
However, the collisional broadening will not 
reduce the total dilepton yield.
Adding the collisional broadening width in the denominator
of Eq. (\ref{form}) requires the inclusion of 
dilepton production also from processes responsible for the collisional
broadening. The net effect is thus not a reduction of the total dilepton
yield but only a broadening of the invariant mass spectrum. Since the
mass resolution in present experiments is comparable to the collisional 
broadening widths, the final results after being corrected by the
experimental resolution will be similar whether we include 
the collisional broadening widths or not. We have thus neglected the vector
meson collisional widths in the present study.
We note that the collisional broadening widths of vector mesons
are reduced when medium-dependent masses are used \cite{haglin}.
 
\subsection{Dilepton emission from the expanding fire-cylinder}
 
In our model, dileptons are emitted not only during 
the expansion of the fire-cylinder but also at the freeze-out.
Let us illustrate the way dilepton production is treated
in the transport model by considering rho meson decay.
Denoting the differential multiplicity of neutral rho mesons
at time $t$ by $dN_{\rho ^0}(t)/dM$, then the differential
dilepton production probability is given by
\begin{eqnarray}\label{sum}
{dN_{e^+e^-}\over dM} =\int _0^{t_f} {dN_{\rho^0} (t)\over dM} 
\Gamma _{\rho^0\rightarrow e^+e^-}
(M) dt + {dN_{\rho ^0}(t_f)\over dM} 
{\Gamma _{\rho^0\rightarrow e^+e^-} (M)\over\Gamma _\rho (M)},
\end{eqnarray}
where the freeze-out time $t_f$ is taken to be 20 fm when
there is little interaction among the constituents of the
fire-cylinder. For the rho meson, whose lifetime is about 1-2 fm/c,
the contribution from the first term in Eq. (\ref{sum}), i.e, 
before the freeze-out, is much more significant than those from the second
term, i.e., after the freeze-out. This remains to be so in the
case of dropping rho meson mass. Although the rho meson lepton decay
width is reduced when its mass decreases in medium (see Eq. (\ref{ree})), 
the increase in the number of low-mass rho mesons in medium is more 
significant, as shown in Fig. 6, leading to an enhanced production of 
low-mass dileptons compared to the free mass case.  The in-medium properties 
of rho meson can thus be studied from the shape of its associated dilepton 
spectrum. 
 
Similar expressions as Eq. (\ref{sum}) 
are used for dilepton yields from other sources.
For omega and phi mesons, whose
lifetimes are more or less comparable to the lifetime of the fire-cylinder,
the determination of their in-medium properties from the shape
of their associated dilepton spectra becomes difficult (see Figs. 13
and 14 below and also Ref. \cite{li95a}). One could, however,
observe the interesting phenomenon of rho-omega splitting in the dilepton
spectra if their masses are reduced in hot dense matter as
also mentioned in Ref. \cite{hoff94}. 
 
Similarly, the $a_1$ Dalitz decay mainly contributes during the 
evolution of the system due to its large width, while the $\omega$ 
Dalitz decay has comparable contributions during the evolution 
and at the freeze-out.  On the other hand, for the Dalitz decay of
$\pi ^0$, $\eta$, and $\eta^\prime$, whose lifetimes are much longer than
the lifetime of the fire-cylinder, contributions from
the first term in Eq. (\ref{sum}), i.e., before the freeze-out, are
negligibly small (see Fig. 18 below). This immediately implies several 
important observations. First of all, there is almost no possibility
to observe the medium modification of, e.g., $\eta$ and/or $\eta^\prime$
meson, from the shapes of their associated dilepton spectra, as they
decay into dileptons {\it long after} the disintegration of the hot dense
matter, when their properties must have returned to those in free space.
This has also been realized in Ref. \cite{kapu95}. Secondly, 
the dilepton yield from the Dalitz decay of $\pi^0$, $\eta$ and $\eta^\prime$
is essentially determined by their abundance at the freeze-out.
Thus, they can be accurately extracted 
once the abundance of these mesons are reliably determined 
experimentally from, e.g., two-photon reconstruction. Finally, any
theoretical calculation (or speculation) of 
their contributions to dilepton production have to be constrained
by their measured abundance by, e.g., the WA80 collaboration
\cite{wa80a,wa80b,wa80c}, as discussed at the end of last section. 
We will come back to this point later.
 
\subsection{Results and comparison with the CERES data}
 
In this subsection, we discuss dilepton production in central
S+Au collisions as obtained in our
calculation. We will show the dilepton invariant mass spectra with
and without the medium modification of vector meson masses, and
with and without the experimental acceptance correction. A mass resolution of
about 10\% of the dilepton invariant 
mass, in accordance with the CERES experiment,
has always been included.
In Fig. 11, the dilepton invariant mass spectra from the Dalitz decay of
$\pi^0$, $\eta$, $\omega$, $\eta^\prime$ and $a_1$, 
as well as the sum of these contributions (solid curve), are shown.
It is seen that for dileptons with mass
below 100 MeV $\pi^0$ Dalitz decay dominates. In the mass region
from 150 to 400 MeV, the eta Dalitz decay is the most important, while
omega Dalitz decay dominates the Dalitz background from 500 to 650 MeV.
The contributions from $\eta^\prime$ and $a_1$ Dalitz decay 
become important only above 650 MeV, where the contributions from
direct decay of vector mesons are much more significant. 
We do not see much difference between the results with and without
the medium modification of meson properties.
Although more mesons are present because of reduced masses in the case where
medium dependence is taken into account, the chemical potential
enhances their numbers in the free mass case.
The dilepton invariant 
mass spectra after applying the experimental acceptance
cut are shown in Fig. 12. 
Note that both the $\eta$ and $\omega$ Dalitz decay contributions in our
calculation are somewhat larger than those in the `cocktail' of the CERES
collaboration \cite{ceres95}, as we have larger $\eta /\pi^0$ and
$\omega /\pi^0$ ratios than what were assumed in Ref. \cite{ceres95}
based on proton-proton and proton-nucleus data
due to their enhanced production in heavy-ion collisions.
 
The dilepton invariant mass spectra from the decay of
rho (including rho mesons formed from pion-pion annihilation,
from the decay of $a_1$ mesons and higher baryon resonances, as well
as primary rho mesons)
and omega mesons are shown in Fig. 13. The dotted curves
are obtained using free vector meson masses, while the solid curves are
obtained using in-medium masses. In the low-mass region, 
the enhancement is about a factor of 3-5 and
is similar to that found in Ref. \cite{li95a} for
heavy-ion collisions at SIS/GSI energies and in Ref. \cite{hoff94}
at AGS/BNL energies. This is mainly from the decay of 
low-mass primary rho mesons in the initial hot dense 
fire-cylinder. We note that as the omega
meson has a longer lifetime it decays mainly at the freeze-out where
its mass has returned to that in free space, leading to the peak
around $m_\omega=783$ MeV in the dilepton invariant mess spectrum.
A similar plot after the
experimental acceptance cut is shown in Fig. 14.
 
The sum of dilepton invariant mass spectra from Dalitz decay
and direct vector meson (including also phi meson)
decay from Figs. 11 and 13 (without experimental acceptance cut)
are shown in Fig. 15. The dotted curve
is obtained with free meson masses, while the solid curve
is obtained with in-medium masses. It is seen that in the mass region
from 2$m_\pi$ to about 550 MeV, there is an enhancement of about a factor
of 3 when the in-medium vector meson masses are used.
The enhancement at low masses is smaller than shown in Figs. 13 and 14
as at these masses the Dalitz decay contribution, which is very similar
in the two cases, becomes important.
 
The comparison with the experimental data is shown in Fig. 16.
The results using the free meson masses are shown
with the dotted curve. It is seen that this model underestimates the
data from 250 to 550 MeV by about a factor of 3, while
it slightly overestimates the data around $m_{\rho ,\omega}$.
Most importantly, the shape of the dilepton spectrum is in clear 
disagreement with the experimental data.
The results of our calculation using in-medium 
meson masses are shown in the figure by the solid curve.
The agreement with the experimental data is greatly improved as a result 
of the factor of 3-4 enhancement in the yield of
dileptons with masses from 250 MeV to about 550 MeV, and some
reduction around $m_{\rho ,\omega}$.
 
\subsection{Comparison with other models}
 
In addition to our calculation using the relativistic transport model,
several different dynamical models have recently been used to 
study dilepton production in heavy-ion collisions at CERN/SPS
energies. In Ref. \cite{gale95} a hydrodynamical model, with the assumption
of quark-gluon-plasma formation and a
first-order phase transition, has been used
for the expansion of the fire-cylinder. The results for the dilepton
invariant mass spectrum are reproduced in
Fig. 17 by the dotted curve (the outstanding peak around 1 GeV would disappear
if an experimental mass resolution of about 100 MeV were properly included
in the calculation of Ref. \cite{gale95}), 
together with our results using free meson masses
(solid curve). In Ref. \cite{cass95}, the string dynamics for the initial
stage of heavy-ion collisions has been combined with
the hadronic transport model to study dilepton production.
The results of Ref. \cite{cass95} using free meson
masses are reproduced in Fig. 17 by the dashed curve. It is interesting to
see that the dilepton spectra obtained in all three calculations
based on different dynamical models agree with each other within 
about 20\%, and all underestimate the data at 
the low-mass region by about a factor of 3 and 
overestimate the data around the rho-meson mass.
 
The failure of the three calculations using free vector meson masses
and the quantitative explanation of the data with in-medium vector
meson masses suggest that 
the observed enhancement of low-mass dileptons by the CERES collaboration
might be the first direct experimental evidence 
for the decrease of vector-meson masses in hot and
dense hadronic matter as a result of partial restoration of chiral symmetry.
This is indeed a significant conclusion and one may ask whether there
are `conventional' or `alternative' mechanisms to explain the data.
 
Conventionally, one might think of the medium modification of
the rho meson spectral function (or equivalently pion electromagnetic
form factor) as a typical nuclear many-body effect. 
It has been shown in Refs. \cite{qiu,herr93} that the rho-meson
spectral function is modified in dense nuclear matter as a result of
the softened pion dispersion relation due to delta-particle$-$nucleon-hole
($\Delta N^{-1}$)
polarization.  The form factor as determined in Ref. \cite{herr93}
has been used in Ref. \cite{cass95} to study its effects on the 
dilepton spectrum. Indeed, one sees the enhancement of low-mass dileptons
as compared to the case with the free-space rho-meson spectral function.
A similar result has also been obtained in Ref. \cite{rapp} in which
not only the delta-particle$-$nucleon-hole polarization but also the 
nucleon-particle$-$nucleon-hole polarization 
in nuclear matter at finite temperature has been included. 
However, the effects found in Refs. \cite{cass95,rapp}, 
though considerably smaller
than the effects from the decreasing rho meson mass, are most likely 
overestimated. Firstly, in heavy-ion collisions at SPS/CERN energies,
the system is highly excited, so it consists not only of nucleons
but also of baryon resonances, especially $\Delta (1232)$. The softening
of the pion dispersion relation due to the delta-particle$-$nucleon-hole
polarization is largely cancelled by the counter-effects from
the nucleon-particle$-$delta-hole polarization. In this case the
softening of the pion dispersion relation depends on an effective density
$\rho _{eff}=\rho _N-{1\over 4}\rho _{\Delta}$ as shown in Ref. \cite{henn94}.
At an initial temperature of 165 MeV and initial baryon density
of 2.5$\rho _0$, we find that $\rho_N$ and $\rho _{\Delta}$ are
about 0.8$\rho_0$ and 0.9$\rho_0$, respectively, 
thus the effective density $\rho_{eff}\approx 0.6\rho_0$.
At this density the change of the rho spectral function is already
insignificant \cite{qiu,herr93}. Including the counter-effects from higher
resonances such as $N^*(1440)$, the effective density will be further reduced.
Moreover, pions outnumber nucleons by about a factor of 5-10
in heavy-ion collisions at CERN/SPS energies, so not all the pions
can be converted into ($\Delta N^{-1}$)'s.  In the calculation
of Refs. \cite{qiu,herr93}, the pion dispersion relation
has been obtained always by putting only one pion into an infinite nuclear
matter.  Also, the mismatch in the pion and nucleon rapidity 
distributions in the initial stage when dileptons are produced
would reduce the effect of delta-hole polarization.
For heavy-ion collisions at SPS/CERN energies, 
with a correct pion dispersion relation as a function of 
not only nucleon but also baryon resonance and pion densities
as well as a proper consideration of the nucleon momentum distribution,
this nuclear many-body effect is expected to be much smaller than 
that found in Ref. \cite{cass95,rapp}.
 
The pion dispersion relation is also softened at finite 
temperature due to coupling to the pion-rho loop \cite{songb}.
The enhancement of low-mass dileptons from this effect is, however, 
largely offset by the reduction in the pion electromagnetic form 
factor at finite temperature \cite{songc}.
 
Alternatively, 
in connection with the interesting possibility of $U(1)_A$ symmetry
restoration that leads to medium modifications
of the $\eta$ and/or $\eta^\prime$ meson masses, 
several speculations have been put forward \cite{huang95,kapu95}.
Looking at the Dalitz decay background, one might suggest that the
medium modification of the $\eta$ and/or $\eta^\prime$ meson masses might
enhance low-meson dileptons.  However, as mentioned previously, the Dalitz
decays of the $\eta$ and $\eta^\prime$ mesons occur dominantly long after the
freeze-out of the system. This is quantified in Fig. 18
where the dilepton spectrum from the decay of eta mesons up to 20 fm/c is
shown by the dotted curve, while their total contribution
is shown by the solid curve. It is clearly seen that 
their contribution to the dilepton yield before the freeze-out, 
i.e., in hot and dense medium where $\eta$ and $\eta^\prime$
masses might be modified, is negligible as compared to that
after the freeze-out when their masses are the same as in free space.
Therefore the shape of the dilepton spectra from the Dalitz 
decay of the $\eta$ and $\eta^\prime$ mesons remains essentially 
the same, be there medium modifications
or not. On the other hand, one might argue that the reduced $\eta$ and
$\eta^\prime$ meson masses increase their production cross section 
and thus increase overall the dileptons from their Dalitz decay. 
However, one needs to keep in
mind an important constraint of the model as provided by the
experimental data on the $\eta/\pi^0$ ratio \cite{wa80a,wa80b,wa80c}.
Since we do reproduce the $\eta/\pi^0$ ratio reasonably well (see
Fig. 9),  we believe that our results for the Dalitz decay
of the $\eta$ and $\eta^\prime$ are quantitatively correct. 
As pointed out earlier, there is
already some enhancement of Dalitz background with respect to 
the `cocktail' of Ref. \cite{ceres95}.
 
In Ref. \cite{huang95}, partial $U(1)_A$ restoration has been studied,
and both the $\eta$ and $\eta^\prime$ meson masses are found to 
decrease at high temperature. It has been found in Ref. \cite{huang95}
that in order to describe quantitatively the $\eta /\pi^0$ ratio
in central S+Au collisions as measured by the WA80 collaboration
\cite{wa80a,wa80b,wa80c}, one needs a finite $\eta$ chemical potential
and a reduced $\eta$ meson mass. This is in agreement with our findings.
We need either a large $\eta$ and $\eta^\prime$ chemical potential,
or a moderate chemical potential with reduced $\eta$ and $\eta^\prime$
meson masses. The enhanced $\eta$ yield in central heavy-ion collisions
as compared to that in proton-proton and proton-nucleus interaction are,
however, not enough to explain the enhanced low-mass dilepton yield 
in the CERES experiments.
 
Similarly, $U(1)_A$ symmetry restoration in hot hadronic matter has
been studied in Ref. \cite{kapu95}, in which the
medium modification  of $\eta^\prime$
mesons and their enhanced production are emphasized.
By increasing the $\eta^\prime$ contribution in the 
`cocktail' of Ref. \cite{ceres95} by 16 times, the authors of Ref. \cite{kapu95}
found that the Dalitz decays, together with hadron-hadron (especially
pion-pion) contributions, might be able to explain the enhancement of low-mass
dileptons in the CERES experiments.
It should, however, be emphasized that
the $\eta^\prime$ contribution in the `cocktail' of Ref. \cite{ceres95} 
has been obtained with an $\eta^\prime /\pi^0$ ratio of about 0.04-0.05
\cite{ceres95,lebc}.
Thus a 16-fold increase of $\eta^\prime$ abundance implies an
$\eta^\prime /\pi^0$ ratio of about 0.7.  Together with
a similar $\eta /\pi^0$ ratio, this gives rise to a final $\eta /\pi^0$ 
ratio of greater than 1. This grossly overestimates the
measured ratio in the most central S+Au collisions by the WA80 collaboration
and is more so for less central CERES experiments. 
 
\subsection{Sensitivity of dilepton spectra to initial conditions}
 
The main uncertainty in our model is the initial conditions
for the meson chemical composition, the transverse momentum distributions
of hadrons, and the rapidity field.
Microscopic models
based on string fragmentations can in principle provide the necessary
information \cite{li96}.  In the present work, only general
features of the results from these models have been used. 
Values of the parameters in the model,
particularly the initial meson compositions, are determined
by fitting the measured proton and pion rapidity and
transverse momentum distributions.
With free vector meson masses, these initial conditions fail to explain
the observed enhancement of low-mass dileptons. 
One may wonder if different initial meson
compositions can be used to increase the dilepton yield
without invoking any medium effects.  Fig. 17 has already 
demonstrated that using an initial meson composition from either string
fragmentation or in chemical equilibrium can not lead to an enhanced
production of low-mass dileptons. 
Here, we will show the results from a calculation
with an initial meson chemical composition that is determined by
the spin-isospin degeneracy 
as in Ref. \cite{bert91}, i.e., the initial ratio of 
the $\pi$, $\rho$, $\omega$, and $a_1$ abundance is 1:3:1:3.
To reproduce the observed pion multiplicity then requires that 
the numbers are 21, 63, 21, 63 for $\pi$, $\rho$, $\omega$, and $a_1$,
respectively. The dilepton invariant mass spectrum 
from this scenario is shown
in Fig. 19 together with the results using the chemical equilibrium
scenario. The dilepton spectra in both calculations 
are essentially the same, with the peak around
$m_{\rho ,\omega}$ slightly lowered and a small enhancement around 300
MeV to 600 MeV in the spin-isospin scenario.
 
We have also studied the dependence of the dilepton invariant mass
spectrum on the parameters $y_L$ and $\sigma$, which characterize the
initial meson rapidity distributions. By varying their values but still
requiring that the final pion rapidity distribution agrees reasonably
with the measured one, we have not found any appreciable change in the 
dilepton invariant mass spectra for both free and in-medium masses.
 
\subsection{Comparision with the HELIOS-3 data}
 
Enhancement of dilepton yields in heavy-ion collisions
in both low-mass and intermediate-mass
regions have also been observed by the HELIOS-3 collaboration
which have measured dimuon invariant mass spectra in both
S+W and p+W collisions at 200 GeV/nucleon \cite{helios95}. 
The enhancement has been
demonstrated in terms of the difference in dilepton
yield per charged particle between S+W and p+W collisions as 
shown in Ref. \cite{helios95}. The data then 
show that the enhancement
is most pronounced around $M\approx 0.5-0.6$ GeV, which provides another
possible indication, independent of the CERES data, 
that the vector (rho) meson mass might be reduced in hot and dense matter.
 
To  see whether our model, that has explained the enhancement of dilepton
yield as observed by the CERES collaboration,  can also explain the
enhancement observed by the HELIOS-3 collaboration in the low-mass region,
we have carried out two calculations for dimuon production. One assumes
free meson masses and the other uses in-medium masses with the same 
parameters as specified earlier. Since the HELIOS-3 collaboration
measures dimuons, with $m_\mu\approx 0.106$ GeV, we need to include
a phase factor of 
$\big(1-(2m_\mu /M)^2\big)^{1/2}(1+2m_\mu^2/M^2)$ 
in Eq. (\ref{opee}) for vector meson decays. The results  are shown in
Fig. 20, and compared with the data from Ref. \cite{helios95}.  Note
that the data are the difference between dilepton yields per charged
particle in S+W and p+W collisions, i.e., 
$\big(\Delta N_{\mu\mu}/N_{ch}\big)_{SW}-\big(\Delta
N_{\mu\mu}/N_{ch}\big)_{pW}$, while our theoretical results include 
only dileptons from the decay of rho and phi mesons 
formed from pion-pion and kaon-antikaon 
annihilation, which are absent in p+W collisions.
The y-axis, $\Delta N_{\mu\mu}/N_{ch} ~(50 {\rm MeV}/{\rm c}^2)^{-1}$, 
represents the total number of dimuon pairs in a mass bin of 50 MeV,
including the experimental acceptance cuts and 
normalized to the total charged-particle
multiplicity in the pseudorapidity region $3.7<\eta <5.2.$
It is seen that with free rho meson mass, the shape of 
the pion-pion contribution is in clear disagreement with the data;
in the mass region below 650 MeV it underestimates the data by
about a factor of 2-3, while around $m_\rho$ it slightly overestimates
the data. This is very similar to the case for the CERES data.
Including decreasing rho-meson mass in hot and dense matter,
the agreement with the data is again greatly improved. Specifically, we have
about a factor of 2 enhancement in the mass region below about 600 MeV
as compared with the case using the free rho-meson mass.  There is also 
some reduction around $m_\rho$ due to the shift of rho-meson peak to
a lower mass. Overall, we have reasonably agreement with the experimental
data up to about 800 MeV. In the mass region around 1 GeV, enhanced dilepton
production in S+W collisions has also been observed which is
most likely due to the kaon-antikaon annihilation that proceeds through
the phi meson \cite{li95a}. 
 
In both experiments \cite{helios95,ceres95,tserruya} 
and our theoretical calculation, 
the enhancement of dilepton yield in the
low-mass region in Fig. 20 is somewhat smaller than that in Fig. 16. 
There are two possible reasons. Firstly, the CERES collaboration measures
dileptons in the mid-rapidity region where initially there are
many low-mass rho mesons, while
the HELIOS-3 collaboration measures the forward-rapidity region with 
a smaller number of low-mass rho mesons. Secondly, in our model the 
dileptons measured in the 
CERES experiments are produced early in the expansion when the
baryon density is high, and the rho meson mass is small.
In the HELIOS-3 experiment, the dileptons measured in the forward
rapidity are produced later in the expansion when interactions lead to a
broader rapidity distribution. The baryon density at which
large rapidity dileptons are produced is thus lower, and
the reduction of the rho meson mass is smaller.
 
\section{general discussions}
 
There have been discussions of processes which might fill in the low-mass
dilepton spectrum. One example is the pionic bremsstrahlung process 
shown in Fig. 21. This process would, indeed, contribute to low-mass 
dileptons.
If the lower left $\rho$-meson is produced by a $\gamma$-ray, one can see 
that this is just the electromagnetic polarizability of the pion, with
the final $\gamma$-ray going into an $e^+e^-$ pair. It is known, however, 
from the Das, Mathur and Okubo relation \cite{das67} that the $\rho$-meson 
does not enter into the polarizability of the pion, although it is most 
important in the pion electromagnetic form factor. A double-$\rho$-seagull
enters with equal magnitude and opposite sign in order to cancel the
process of Fig. 21. This means that the $\rho$ and $\pi$ couple
uniquely through the $a_1$ as intermediate state \cite{hols90}, and
this has been employed in our study.
 
For systems with nonzero charge, the $0^{++}$ sigma meson can couple 
through $\pi^+\pi^-$ to a single virtual photon and dileptons
\cite{weldon92}.  This bears some relation to the known possibility
in Walecka theory at finite density in nuclei that the sigma can couple to 
the time component of the $\omega$-meson, which can then decay into 
dileptons.  Thus, at finite density (and breaking of Lorentz invariance) 
the sigma cannot decay into a single photon or dilepton pair does not 
hold. Weldon \cite{weldon92}
makes the argument that the sigma mass will drop with density and/or 
temperature, so that dileptons from the sigma decay through a single virtual
photon will fill up the low-mass region of dileptons. 
However, as long as the vector mesons
are made in sufficient number, as we find for the process
$\pi\pi\leftrightarrow\rho$, it does not matter how many additional processes
are introduced. Very simply, an equilibrium distribution of vector mesons, 
with appropriate Boltzmann factors, is formed. If additional processes are
added to make more vector mesons, then the reverse processes will remove it.
 
The goodness of our initial conditions, mainly the hadron
momentum distributions, was found by Cassing {\it et al.}
\cite{cass95} and in the RQMD \cite{li96}. 
These authors began from initial distributions generated
from string dynamics without the assumption of specific 
functional forms for the transverse momentum spectra and rapidity 
distributions. Yet it can be seen from our Fig. 17 
(and a similar figure from Ref. \cite{li96}) that their results
are similar to ours and to the hydrodynamical ones of Srivastava 
{\it et al.} \cite{gale95}. Our somewhat larger contribution at
low masses results from our finite pion and eta chemical potentials. 
 
Finally, one can question our use of Walecka theory, although many people
have used it at constituent quark level. It is true that the constituent
quark effective masses in Walecka theory go to zero only as $\rho\to\infty$,
whereas we believe Nambu$-$Jona-Lasinio theory to be more correct in letting 
the masses go to zero at finite (and not too high) densities.  We have, 
however, started the evolution of our fire-cylinder when the constituent
quark has about one-third of its on-shell mass, and the difference 
between Walecka theory and Nambu-Jona-Lasinio theory should not be 
too great here. 
It can be shown \cite{buballa} that the field energy in the Walecka theory 
acts like a bag constant, and that the parameters in the Walecka linear
sigma-omega model are such that if $m_Q^*$ is brought to zero, this effective
bag constant faithfully reproduces that from lattice gauge calculations
in QCD.  The latter is obtained from changes in quark/gluon condensates as
the temperature goes through the chiral transition \cite{koch93}.
In our expansion of the fire-cylinder it is important to have the correct value 
for the bag constant.  The energy going into the bag constant takes away
from the energy in expansion, providing negative pressure, so the 
expansion is slower as can be seen from our Fig. 6.
 
As noted above, RQMD-type string dynamics calculations have been and 
are being carried out to
establish initial conditions.  In our dropping mass scenario, at the chiral 
phase transition the vector meson masses are zero. Thus, there is a period
between the phase transition and the time we begin evolution of the 
fire-cylinder in which the masses are smaller than those in our calculation.
The decay of these very low mass vector mesons should not be very important,
because of the small phase space and because they will chiefly give
dileptons in the low-mass region, which is dominated by the background
of Dalitz decays.  Work is, however, underway to propagate the masses
described in Nambu$-$Jona-Lasinio theory. This theory does bring the
constituent quark masses smoothly to zero, and the bag constant can be
arranged to match the decondensation energy obtained in lattice gauge 
calculations.
 
\section{summary}
In conclusion, using the relativistic transport model for the expansion stage
of the fire-cylinder formed in the initial stage of heavy-ion collisions
at SPS/CERN energies, we have studied the enhancement of low-mass dileptons
in central S+Au and S+W collisions as recently observed 
by the CERES and the HELIOS-3 collaboration.
The initial conditions for the fire-cylinder are constrained by
microscopic simulations based on string dynamics as well as 
experimentally measured proton and pion rapidity distributions and
spectra. We have 
included the Dalitz decay of $\pi ^0$, $\eta$, $\eta^\prime$, $\omega$
and $a_1$ mesons, the direct decay of primary 
$\rho ^0$, $\omega$ and $\phi$ mesons, and the
pion-pion annihilation that proceeds through the $\rho^0$ meson,
the pion-rho annihilation that proceeds through the $a_1$ meson, 
and the kaon-antikaon annihilation that proceeds through the $\phi$ meson.
We have found that the modification of vector meson
properties, especially the decrease of their mass due to the partial
restoration of the chiral symmetry, in hot and dense hadronic
matter, provides a quantitative explanation
of the recently observed enhancement of low-mass dileptons
by the CERES collaboration in central S+Au collisions and by the HELIOS-3
collaboration in central S+W collisions. In comparison with the latter, we 
have, however, too few phi mesons. It has been shown in Ref. \cite{ko94b}
that phi meson mass is considerably brought down by the substantial
(positive) $\langle \bar ss\rangle $ present in the fire-cylinder. 
We are unable, as of yet, 
to include this in a thermodynamically consistent way in the transport
model.  
 
To continue this study, we have already started to carry out 
a calculation \cite{li96} in which the initial conditions are 
completely given by 
the string dynamics. Although a similar study has previously been
carried out by Cassing {\it et al.} \cite{cass95,cassing}, our approach
differs from theirs in the treatment of vector meson masses in hot
dense matter as discussed in Section II. Preliminary results
show that the results presented in the present paper remain essentially
unchanged. This more microscopic approach allows us to make
predictions regarding low-mass dilepton production from collisions
between heavier nuclei, where the medium effects are expected to
be more appreciable. 
 
\vskip 1cm
We are grateful to Peter Braun-Munzinger, Wolfgang Cassing, Kevin Haglin,
Volker Koch, Zheng Huang, 
Mannque Rho, Bikash Sinha, Edward Shuryak, and Chungsik Song 
for useful discussions.
Also, we thank Alex Drees, Itzhak
Tserruya and Thomas Ullrich for communications about the CERES
experiments, Ivan Kralik for communications about the HELIOS-3
experiments, Michael Murray for communications about the NA44 
experiments, and Frank Plasil and Karl-Heinz Kampert
for communications about the
WA80 experiments. GQL and CMK are supported by the
National Science Foundation under Grant No. PHY-9212209 and 
No. PHY-9509266, and GEB is supported by the
Department of Energy under Grant No. DE-FG02-88ER40388.

\pagebreak
 
\centerline{\bf Figure Captions}
\begin{description}
 
\item {Fig. 1:} In-medium mass of nucleon (left panel) and rho-meson
(right panel) as a function of temperature at several baryon densities.
 
\item {Fig. 2:} Initial proton and pion transverse momentum and rapidity 
distributions.
 
\item {Fig. 3:} Baryon (left panel) and meson (right panel) distributions
in x-z plane at several time steps.
 
\item {Fig. 4:} Longitudinal (left panel) and transverse (right panel)
flow velocities at freeze-out.
 
\item {Fig. 5:} Time evolution of the nucleon, $\Delta (1232)$ and higher
baryon resonance abundance.
 
\item {Fig. 6:} Time evolution of the meson abundance.
 
\item {Fig. 7:} Time evolution of the central baryon density.
 
\item {Fig. 8:} Proton and pion transverse momentum spectra.
Dotted and solid histograms are obtained from simulations
based on the relativistic transport model
with free and in-medium vector meson
masses, respectively. Open and solid circles are
experimental data from the CERES \cite{ceres94} and the WA80
\cite{wa80a} collaboration, respectively.
 
\item {Fig. 9:} Same as Fig. 8 for proton and pion rapidity distributions.
The experimental data from the NA35 \cite{na35b}
are given by open circles, while open squares
are obtained by reflecting the data with respect to $y\approx 2.65$.
Solid circles are the experimental data from the NA44 collaboration
\cite{na44}.
 
\item {Fig. 10:} $\eta/\pi^0$ ratio.
Open circles and squares are obtained from simulations
based on the relativistic transport model
with free and in-medium vector meson
masses, respectively. The solid circles and squares are
experimental data from the WA80 collaboration for
the minimum-biased \cite{wa80c} and central \cite{wa80b} S+Au collisions, 
respectively.
 
\item {Fig. 11:} Dilepton invariant
mass spectra from Dalitz decay using free masses
(left panel) and in-medium masses (right panel).
 
\item {Fig. 12:} Same as Fig. 11 with the experimental
acceptance cut.
 
\item {Fig. 13:} Dilepton invariant
mass spectra from the direct decay of rho and omega
mesons, using free masses (dotted curves) and in-medium masses
(solid curves).
 
\item {Fig. 14:} Same as Fig. 13 with the experimental
acceptance cut.
 
\item {Fig. 15:} Dilepton invariant
mass spectra including both Dalitz decay and direct decay
of vector mesons. The dotted and the solid curve are obtained
using the free and in-medium masses, respectively.
 
\item {Fig. 16:} Same as Fig. 15 with the experimental
acceptance cut. 
The experimental data from the CERES collaboration \cite{ceres95} are
shown by the solid circles, with the statistical errors given by
bars. The brackets represent the square root of the quadratic 
sums of the systematic and statistical errors. (We have followed
the request of Itzhak Tserruya of the CERES collaboration to show
the experimental errors in this way instead the linear sum 
used in the original publications \cite{ceres95}, which were often 
misinterpreted.)
 
\item {Fig. 17:} Comparison of different calculations with free meson masses.
The solid curve is from the present work, while the dotted and dashed curves
are from Ref. \cite{gale95} and Ref. \cite{cass95}, respectively.
 
\item {Fig. 18:} Dilepton invariant mass spectra from $\eta$ Dalitz decay.
The dotted curve gives the spectrum up to 20 fm/c, while the solid
curve is the total. 
 
\item {Fig. 19:} Dilepton invariant 
mass spectra obtained using free meson masses.
The dotted and the solid curve are obtained, respectively, using the 
spin-isospin scenario
and the chemical equilibrium scenario for the initial meson 
composition.
 
\item {Fig. 20:} Dimuon mass spectra from pion-pion annihilation.
The solid and dotted histograms are obtained with in-medium and free
meson masses, respectively. Solid circles are the experimental data 
for the difference 
in the dimuon yields per charged particle
between the S+W and p+W the collision from the HELIOS-3 collaboration
\cite{helios95}.
 
\item{Fig. 21:} Pionic bremsstrahlung.
 
\end{description}
 
\end{document}